\documentstyle{article}
\newcommand{\be}{\begin{equation}}
\newcommand{\ee}{\end{equation}}
\newcount\sectionnumber
\def\sect
{\global\equationnumber=0
\global\advance\sectionnumber by 1
\the\sectionnumber . }
\newcount\equationnumber
\def \num
{\eqno{\global\advance\equationnumber by 1
\left(\the\sectionnumber .\the\equationnumber \right)}}

\def\boxit#1{\vbox{\hrule\hrule\hbox{\vrule\vrule\kern3pt\vbox{\kern3pt#1\kern3pt}\kern3pt\vrule\vrule}\hrule\hrule}}

\def\comma{ \hspace{2mm}, }
\def\period{ \hspace{2mm}. }
\def\s{ \hspace{1mm} }
\def\b{ \hspace{2mm} }
\def\B{ \hspace{4mm} }
\def\EXP{ {\rm e} }
\def\Del{ \nabla }
\def\di{ \partial }
\def\de{ {\rm d} }

\def\Rtwo{ \sqrt{2} }
\def\minus{ \mbox - \, }
\def\const{ 8\,\pi\,G }

\begin{document}
\onecolumn
\begin{flushright}
  WATPHYS TH-94/07\\
  gr-qc/yymmddd
\end{flushright}
\vfill
\begin{center}
  {\Large \bf Inside a Spinning Black String} \\
  \vfill
  J.S.F. Chan$^{(1)}$ and R.B. Mann$^{(1,2,3)}$ \\
  \vspace{2cm}
  (1) Department of Applied Mathematics, University of Waterloo,
  Waterloo, Ontario, Canada, N2L 3G1 \\
  (2) Department of Physics, University of Waterloo,
  Waterloo, Ontario, Canada N2L 3G1 \\
  (3) Department of Applied Mathematics and Theoretical Physics,
  Cambridge University, Cambridge, U.K. CB2 9EW\\
  \vspace{2cm}
  PACS numbers: 97.60Lf,04.70.-s,04.20.Dw \\
  \vfill

  \begin{abstract}
    We show that mass inflation occurs inside spinning black cosmic
    string, which is a solution of a low-energy effective string theory
    in $(3+1)$-dimensions. This confirms Poisson and Israel's conjecture
    that the inner mass parameter diverges even if spacetime is not
    spherically symmetric.
  \end{abstract}

  \vfill
\end{center}
\clearpage


The interior structure of black holes has become a subject of
increasing interest in the past few years. While it has been
demonstrated within general relativity that, given the hypothesis
of cosmic censorship, a collapsing body has an exterior spacetime
which relaxes to that of a Kerr-Newman (KN) black hole (with radiative
perturbations decaying as advanced time increases according to a power
law), the question of what happens to the infalling matter is much
less well established. Either the infalling matter encounters a region
of diverging spacetime curvature (where the effects of quantum gravity
should dominate) or it avoids this region, emerging into another
universe after having passed through a Cauchy horizon.

Penrose \cite{Penrose} pointed out quite some time ago that the
infinite blueshift of ingoing radiation renders the Cauchy horizon of
the KN solution unstable. However much more recently our understanding
of black hole interiors has improved due to the investigations of
Poisson and Israel \cite{Poisson}, who showed that the mass parameter
inside a Reissner-Nordstr\"{o}m (RN) black hole becomes unbounded due
to the presence of ingoing and backscattered outgoing radiation. This
was subsequently confirmed by Ori \cite{Ori}, who constructed an exact
solution of the Einstein-Maxwell equations that exhibited this
phenomenon in a simpler model in which the outgoing radiation was a
null shell. Since a KN spacetime has a causal structure similar to
that of the RN case, this suggested that the KN Cauchy horizon would
exhibit similar behaviour. However Ori also pointed out that the mass
inflation singularity is so weak that its tidal forces do not
necessarily destroy any physical objects, an issue which has generated
some controversy \cite{Herman,Bonnano} particularly since all models
at that time had assumed spherical symmetry. More recently mass
inflation has also been shown to take place in lower-dimensional
analogs of the RN solution, both in $(1+1)$ \cite{JChan,Droz,Balb}
and $(2+1)$ dimensions \cite{Husain}, and for the charged spinning BTZ
black hole \cite{JKR}.

Here we demonstrate that mass inflation takes place in $(3+1)$
dimensions even in the absence of spherical symmetry. Specifically,
we show that an exact solution to a low-energy effective string theory
which can be interpreted as a rotating black cosmic string
\cite{Kaloper} has a mass parameter which diverges due to the
interaction of ingoing and outgoing radiation.

The effective $(3+1)$-dimensional action we consider describes the
dynamics of the background field formulation of string theory to order
${\cal O}((\alpha^\prime)^0)$ and is given by
$$
  S = \int \de^4 x\,\sqrt{\minus g} \left[ R - (\Del \psi)^2
  - \frac{\EXP^{\minus 2\,\Rtwo\,\psi}}{3}\,H_{\mu \nu \sigma}\,
  H^{\mu \nu \sigma} \right.
$$
\be
  + \left. 2\,\Lambda\,\EXP^{\Rtwo\,\psi}
  - \const\,{\cal L}_M\,\right] \comma \label{Act} 
\ee
where $\psi$ is the dilaton field,
$H_{\mu \nu \sigma} = \di_{[\sigma} B_{\mu \nu]}$ is the curl of the
skew-symmetric Kalb-Ramond field $B_{\mu \nu}$ and $\Lambda$ is a
positive constant. The term ${\cal L}_M$ is the matter Lagrangian for
the infalling radiation which is independent of the dilaton and
Kalb-Ramond fields.

Varying (\ref{Act}) with respect to the fields $\psi$, $B_{\mu \nu}$
and $g_{\mu \nu}$, yields
\be
  \frac{\Del^2 \psi}{\Rtwo} + \frac{H^2}{3}\,\EXP^{\minus 2\,\Rtwo\,\psi}
  + \Lambda\,\EXP^{\Rtwo\,\psi} = 0 \comma \label{FE1} 
\ee
\be
  \Del_\lambda \left[\,\EXP^{\minus 2\,\Rtwo\,\psi}\,H^{\mu \nu
\lambda}\,\right] = 0 \comma
  \label{FE2} 
\ee
\begin{eqnarray}
  \const\,T_{\mu \nu} & = &
  G_{\mu \nu} - \Lambda\,\EXP^{\Rtwo\,\psi}\,g_{\mu \nu} \nonumber \\ &&
  - \s \EXP^{\minus 2\,\Rtwo}\,\left[\,
  H_{\mu \lambda \sigma}\,H_\nu^{\b \lambda \sigma}
  - \frac{H^2}{6}\,g_{\mu \nu}\,\right] \nonumber \\ &&
  - \s \left[\,\Del_\mu \psi\,\Del_\nu \psi
  - \frac{(\,\Del \psi\,)^2}{2}\,g_{\mu \nu}\,\right]
  \period \label{FE3} 
\end{eqnarray}
This set of equations has a static, vacuum solution
\begin{eqnarray}
  \psi & = &
  \frac{1}{\Rtwo}\,\ln\left(\,\frac{\Lambda}{2\,Q^2}\,\right) \comma
  \label{SS1} \\ 
  H_{\mu \nu \sigma} & = &
  \frac{\Lambda^2}{4\,Q^3}\,r\,\epsilon_{4 \mu \nu \sigma} \comma
  \label{SS2} 
\end{eqnarray}
\be
  \de s^2 =
  \minus N^2\,\de t^2 + \frac{\de r^2}{N^2}
  + r^2\,\left(\,N^\phi\,\de t + \de \phi\,\right)^2
  + \de z^2 \comma \label{SS3} 
\ee
\begin{eqnarray}
  N^2 & = & \lambda\,r^2 - M + \frac{J^2}{4\,r^2} \comma
  \label{SS4} \\ 
  N^\phi & = & \minus \frac{J}{2\,r^2} \comma \qquad
  \lambda = \frac{\Lambda^2}{4\,Q^2} \period
  \label{SS5} 
\end{eqnarray}
The solution given here can be interpreted as describing a straight,
spinning black cosmic string of infinite length \cite{Kaloper}.
However the metric in reference \cite{Kaloper} is not a solution of
equations (\ref{FE1})--(\ref{FE3}) except when the $\rho^2_+$ in
\cite{Kaloper} equals $M(4/\lambda -1)$ or $J = 0$, in which case
it is equivalent to the metric (\ref{SS3}) above.

As the coordinate $r$ has a range of $[\,0 \comma \infty\,)$ and
$\phi$ has a period of $2\,\pi$, the spacetime manifold has a
topology $S^1 \times \Re^3$ which represents a static, axially
symmetric spacetime. The metric (\ref{SS3}) is equivalent to
$\de s^2 = \de s^2_{BTZ} + \de z^2$, where $\de s^2_{BTZ}$ is the
BTZ metric for 3-dimensional Einstein theory with cosmological
constant $\lambda$ \cite{BTZ}. Thus we interpret the constants of
integration $M$ and $J$ as the mass and angular momentum per unit
length of the stringy black hole respectively; $Q$ is the charge
of the axion field dual to $H_{\mu \nu \sigma}$ \cite{Kaloper}.

We now consider a solution to (\ref{FE1})--(\ref{FE3}) in the presence
of a cylindrically symmetric cloud of infalling null dust, with
stress-energy-momentum tensor $T_{\mu \nu} = \rho(v)\,l_\mu\,l_\nu / r$.
The term $\rho$, which corresponds to the density of the  dust
with 4-velocity $l_\mu$, must be a function only of $v$ in order to
satisfy the conservation laws. An exact solution to the field
equations is now given by
\be
  \de s^2 =
  \minus \alpha\,\de v^2 + 2\,\de v\,\de r
  - J\,\de v\,\de \theta + r^2\,\de \theta^2 + \de z^2 \comma
  \label{vS} 
\ee
where $\alpha = \lambda\,r^2 - m(v)$, with $m'(v) = 16\,\pi\,G\,\rho(v)$;
the solutions for $\psi$ and $H_{\mu\nu\sigma}$ are still given by
(\ref{SS1}) and (\ref{SS2}) respectively. For constant $m(v)=M$ a
coordinate transformation
\begin{eqnarray*}
  v & := & t + \int^r \frac{\de \zeta}{N^2(\zeta)} \comma \\
  \theta & := &
  \phi - \int^r \frac{N^\phi(\zeta)}{N^2(\zeta)}\,\de \zeta \comma
\end{eqnarray*}
transforms (\ref{vS}) to (\ref{SS3}).

Since one can show that $v$ is a null coordinate, this solution
represents a spinning black string with angular momentum per unit
length $J$ irradiated by a flux of incoming radiation with
synchronized angular momentum. It is straightforward to show that
a null geodesic with no intrinsic spin and constant $z$ moving in
this spacetime obeys the equations
\begin{eqnarray}
  2\,\dot{v}(\omega)\,\dot{r}(\omega) & = &
  N^2\,\dot{v}^2(\omega) \comma \label{NC} \\ 
  \di_r N^2\,\dot{v}^2(\omega) & = &
  \minus 2\,\ddot{v}(\omega) \comma \label{GE} 
\end{eqnarray}
where $\omega$ is an affine parameter for the null curve.

We assume that $J$ is non-zero in order that the spacetime have a
Cauchy horizon. Consider the matching of two patches of the solution
(\ref{vS}) along outgoing null cylinder S. We denote the space
enclosed by S as region II which is characterized by a mass function
$m_{2}(v_2)$. Similarly, we call the complement of II as region I
which has a mass function $m_{1}(v_1)$. The cylindrical boundary is by
$r = R(\omega)$. If we define ${\cal Z}(\omega) := 2\,R(\omega) /
\dot{v}(\omega)$,
it is not difficult to show that the null condition (\ref{NC}) and
geodesic equation (\ref{GE}) yield the matching conditions
\be
  m_{i}(v_{i}(\omega)) =
  {\cal M}(R(\omega)) + R(\omega){\cal M}'(R(\omega))
  - \dot{{\cal Z}}_{i}(\omega) \comma \label{mE} 
\ee
\begin{eqnarray}
  v_{i}(\omega) & = &
  2\,\int^\omega R(\zeta) / {\cal Z}_{i}(\zeta)\,\de \zeta \comma
  \label{vE} \\ 
  {\cal Z}_{i}(\omega) & = &
  R(\omega)\,\left[\,Z_{i} + \int^\omega_0
  {\cal M}'(R(\zeta))\,\de \zeta\,\right] \comma \label{zE} 
\end{eqnarray}
where the function ${\cal M}$ is defined as
\begin{eqnarray}
  {\cal M}(r) & = & \lambda\,r^2 + \frac{J^2}{4\,r^2}
  \b = \b N^2(v,r) + m(v) \B \label{cME} 
\end{eqnarray}
and the subscript $i$ has a value either `1' or `2' to denote
quantities defined in the respective regions. The terms $Z_{i}$ in
equation (\ref{zE}) are integration constants. Without loss of
generality, we assume $\omega$ to be negative in region I and
vanishing at the Cauchy horizon. Given the boundary function $R$,
equations (\ref{mE}) to (\ref{zE}) determine the evolution of the
spacetime.

We define the mass per unit length of the outgoing cylinder S as \cite{Dray}
\begin{equation}
  \Delta m(\omega) := m_{2}(\omega) - m_{1}(\omega) =
  (Z_{1} - Z_{2})\,\dot{R}(\omega) \period \label{MS} 
\end{equation}
Furthermore, we define a constant $M := m_{1}(\omega) + \delta m(\omega)$
as the final mass per unit length of the black string observed in
region I after it has absorbed all the incoming radiation and we
interpret $\delta m$ as the mass per unit length of the flux of
ingoing radiation. Because the Cauchy horizon $R_c$ corresponds to
the limit $v_{1} \rightarrow \infty$, we expect that
\begin{eqnarray*}
  \lim_{\omega \rightarrow 0^{\minus}} \dot{v}_{1}(\omega) =
  \frac{2}{Z_{1}} = \infty
\end{eqnarray*}
implying $Z_{1} = 0$. Since $\dot{R}(\omega)$ is expected to be
negative inside the outer horizon, the sign of $Z_{2}$ must be
positive in order to have a positive-energy cylinder S.

If we define $k_o := \minus {\cal M}'(R_c) / 2$, equation (\ref{zE})
can be written as
${\cal Z}_{i}(\omega) \approx R(\omega)\,( Z_{i} - 2\,k_o\,\omega )$
for small $\omega$. Because of (\ref{cME}), the slope of ${\cal M}(R)$
close to the inner horizon, at which $N^2 = 0$, must be negative no
matter what the precise form of $N^2$ is. Thus $k_o$ is positive
definite for sufficiently small $|\,\omega\,|$. Finally, as equations
(\ref{NC}) and (\ref{vE}) yield small $\omega$ approximations
\begin{eqnarray*}
  2\,\dot{R}(\omega) \approx
  \minus \left| M - m_{1}(v_{1}(\omega)) \right|\,
  \dot{v}_{1}(\omega) \approx \frac{\delta m(\omega)}{k_o\,\omega} \comma \\
  v_{1}(\omega) \approx
  \minus \frac{1}{k_o}\,\ln|\,\omega\,| \quad {\rm and} \quad
  v_{2}(\omega) \approx \frac{2}{Z_{2}}\,\omega \comma
\end{eqnarray*}
we obtain an equation for the inner mass function $m_{2}$ as
\be
  m_{2}(\omega) \approx
  M - h\,\left(\,1 + \frac{1}{k_o\,v_{2}}\,\right)\,k_o^p\,
  \left|\,\ln\left|\frac{Z_2\,v_2}{2}\right|\,\right|^{\minus p}
  \comma \label{m2E} 
\ee
where a power law fall off $\delta m(\omega) \sim h v_1^{-p}$ has been
assumed \cite{Price}. As a result, $m_{2}$ diverges to positive
infinity as $v_2\to 0^{\minus}$, since $Z_{2}$ and $k_o$ are
positive definite quantities but $\omega$ approaches to zero from
below. In other words, the inner mass per unit length of the black
string diverges when the outgoing cylinder S approaches the Cauchy
horizon.

The Ricci and Kret\-sch\-mann scalars for this space\-time
are given by $R = \minus 6\,\lambda$ and
$R_{abcd}\,R^{abcd} = 12\,\lambda^2$ respectively. Thus this
$(3+1)$-dimensional spacetime is radically different from the RN
(and perhaps KN \cite{Bonnano}) spacetime in which the latter
quantity diverges at the Cauchy horizon. Although the Riemann
tensor in comoving tetrad coordinates becomes unbounded due to the
inflating inner mass parameter, one can show that the distortion
due to the tidal force, which is proportional to $R^a_{\b00b}$
where $a$ and $b$ are spatial indices, is finite. This is because
the worst component of $R^a_{\b00b}$ is just
$\minus (\lambda + m'(v) / (2\,r))$; integrating this component
twice gives a bounded distortion at the Cauchy horizon even though
mass parameter diverges there \cite{Ori,Bonnano}. Thus a maximal
extension of this spacetime to other universes does not seem to be
forbidden by an unbounded mass function.

\section*{Acknowledgements}
  We would like to thank Natural Sciences and
  Engineering Research Council of Canada for the support of this work.

\end{document}